\documentclass[preprint,12pt]{elsarticle}

\usepackage{amsmath}
\usepackage{amssymb}
\usepackage{epstopdf}
\usepackage{color}

\begin{document}

\begin{frontmatter}

\title{Exploring optimal institutional incentives for public cooperation}

\author[chn]{Shengxian Wang}
\author[chn]{Xiaojie Chen\corref{cor}}
\ead{xiaojiechen@uestc.edu.cn}
\author[hu]{Attila Szolnoki}

\cortext[cor]{Corresponding author}

\address[chn]{School of Mathematical Sciences, University of Electronic Science and Technology of China, Chengdu 611731, China}
\address[hu]{Institute of Technical Physics and Materials Science, Centre for Energy Research, Hungarian Academy of Sciences, P.O. Box 49, H-1525 Budapest, Hungary}

\begin{abstract}

Prosocial incentive can promote cooperation, but providing incentive is costly. Institutions in human society may prefer to use an incentive strategy which is able to promote cooperation at a reasonable cost. However, thus far few works have explored the optimal institutional incentives which minimize related cost for the benefit of public cooperation. In this work, in combination with optimal control theory we thus formulate two optimal control problems to explore the optimal incentive strategies for institutional reward and punishment respectively. By using the approach of Hamilton-Jacobi-Bellman equation for well-mixed populations, we theoretically obtain the optimal positive and negative incentive strategies with the minimal cumulative cost respectively. Additionally, we provide numerical examples to verify that the obtained optimal incentives allow the dynamical system to reach the desired destination at the lowest cumulative cost in comparison with other given incentive strategies. Furthermore, we find that the optimal punishing strategy is a cheaper way for obtaining an expected cooperation level when it is compared with the optimal rewarding strategy.
\end{abstract}

\begin{keyword}
\texttt Evolutionary game dynamics\sep\ Evolution of cooperation\sep\ Control theory\sep\ Public goods game
\end{keyword}

\end{frontmatter}

\section{Introduction}

Cooperation is of vital importance in the contemporary era and identified as an essential principle of evolution \cite{axelrod06,nowak_s06,rand_tcs13}. However, self-regarding individuals constantly strive to maximize their own personal benefit which would inevitably lead
to the collapse of cooperation, in the scenario where what is best for the collectivity is at odds with what is best for an individual. This is referred to as social dilemma of cooperation \cite{doebeli_el05,szabo_pr07,perc_jrsi13,wang_plr15}, which has received considerable attention in recent years \cite{ohtsuki_n06,masuda_prsb07,fotouhi19,fu_pre09, du_epl09,rong_pre10,helbing_njp10,fu_sr12,han_sr13,zhang_lm_epl19,zhang_sr15,han_jrsi15,szolnoki_pre11b,hintze_pb15,perc_pla16,liu_sr17,perc_epj18,yang_amc18,stojkoski_pre18,wang_amc18,hilbe_pnas18}. To overcome this conflict social institutions frequently apply two control mechanisms, that is, rewards (positive incentives) for cooperation
\cite{gurerk_s06,szolnoki_epl10,sasaki_dga14,chen_fbn14,szolnoki_prx13,chen_jrsi15,sasaki_pnas12,han_srep18,cimpeanu,han_proc18}
and punishments (negative incentives) for defection \cite{pacheco_plr14,hilbe_pnas14,henrich_s06,sigmund_n10,szolnoki_pre11,perc_sr12,vasconcelos_ncc13,szolnoki_prx17,vasconcelos_mmmas15,perc_pr17,chen_pcbi18}.

In theory, cooperative behavior will occur as long as the amount of incentives outweighs the payoff difference between cooperators and defectors no matter whether incentive is positive or negative. However, executing an incentive strategy into the system is costly and the cumulative costs for different incentive strategies could be significantly different. This observation has inspired several studies to explore efficient incentive strategies for the evolution of cooperation at a low cost \cite{chen_fbn14,chen_jrsi15,sasaki_pnas12,han_srep18}. Recently, Sasaki et al. have introduced institutional positive and negative incentives into the public goods dilemma respectively and found that social dilemma of cooperation can be overcome by both institutional punishment and reward. More interestingly, they have found that institutional punishment can overcome social dilemma at a much lower cost with the help of optimal participation \cite{sasaki_pnas12}. Subsequently, Chen et al. have proposed an optimal incentive strategy in combination with institutional punishment and reward, called ``first carrot, then stick", which is able to unexpectedly succeed in promoting cooperation. Furthermore they also demonstrated that the ``first carrot, then stick" strategy makes full cooperation established and recovered at lower cost under wider range of conditions than either reward or punishment alone no matter well-mixed or spatially structured populations are considered \cite{chen_jrsi15}. Nevertheless, it is worth pointing out that these previous works fixed the per capita incentive, which is an important parameter determining the cumulative cost during the evolutionary process, and they focused on the dominant low-cost incentive strategy. In other words, they did not explore the optimal incentive strategy with the minimal cumulative cost for institutional punishment or reward, and thus it remained unclear the specific forms of optimal incentive strategies with the minimal cost for the evolution of cooperation when the per capita incentive is considered as a control variable.

Recently, optimal control theory \cite{evans05,geering07,lenhart07} has been introduced to study evolutionary game dynamics \cite{griffin_pre17,bravetti_sr18,wang_ams18}, which seems to be an alternative approach for studying the optimal incentive strategy with the minimal cost for the evolution of cooperation. For example Griffin and Belmonte have studied the problem of designing an optimal success tax in a three-species public goods game and formulated the problem as an optimal control problem \cite{griffin_pre17}. Furthermore Bravetti and Padilla have introduced an extension of the replicator equation, called the optimal replicator equation, by using optimal control theory \cite{bravetti_sr18}. Additionally Wang et al. have studied the problem of designing reward and compensation control strategies for promoting the evolution of cooperation in the prisoner's dilemma game \cite{wang_ams18}. However, it is worth mentioning that these previous works do not involve the investigation of the optimal incentive strategies with the minimal cumulative cost for the evolution of cooperation, and thus it could be interesting to study the optimal strategies of institutional incentives by using optimal control theory.

In this work, we thus consider the per capita incentive as the control variable in the replicator equation for the public goods game with institutional punishment or reward and aim to explore the optimal control protocols for the control variable in well-mixed populations. Based on optimal control theory, we set a cost function as the objective functional and formulate two optimal control problems for institutional reward and punishment respectively. By using the approach of Hamilton-Jacobin-Bellman equation (HJB equation) \cite{evans05,geering07,lenhart07}, we theoretically obtain the specific optimal control protocols for institutional punishment and reward respectively such that the set objective functional is minimized, i.e., the cumulative cost amount is minimized. We find that the optimal control incentive strategy with the minimal cumulative cost can make the system converge to the homogeneous state of full cooperation. In addition, we provide numerical examples to respectively confirm that the amount of cumulative cost for the optimal control strategy of positive (negative) incentive is lowest when comparing with other given protocols of positive (negative) incentives. We further find that the amount of cumulative cost for the obtained optimal negative incentive strategy is lower than that for the obtained optimal positive incentive strategy in almost all the initial conditions.

\section{Model and methods}

\subsection{Public goods game with dynamic incentives}

Our model is based on the public goods game (PGG) in an infinitely large, well-mixed population, in which each individual can choose to cooperate or defect. In the game, individuals are randomly selected from the population to form a $n$-player group with $n\geq2$. In the formed group, each cooperator contributes the fixed amount of investment $c>0$ to the common pool, while a defector contributes nothing. The total investments to the pool are then multiplied by a synergy factor $r>1$ and allocated equally among all the $n$ group members. The social dilemma arises when $r<n$ and nobody prefers to cooperate if no external incentive mechanism is considered \cite{sasaki_pnas12}.

The basic model can be extended by institutional incentives \cite{chen_jrsi15,sasaki_pnas12}. We assume that apart from the payoffs originated from the PGG, each group receives the total incentives stipulated by an institution in the form $nu$ to be used either for rewarding cooperators or for punishing defectors, where $u \;(u>0)$ is the per capita incentive. If the total incentives are used for rewarding, each cooperator obtains a reward $anu/(n_{C}+1)$, where $n_{C}$ denotes the number of cooperators among the other $n-1$ players and the factor $a$ denotes the leverage of rewarding \cite{chen_jrsi15}. Accordingly, a cooperator receives the payoff
$$
\Pi_{C}=\frac{rc(n_{C}+1)}{n}-c+\frac{anu}{n_{C}+1},
$$
while a defector in the same group receives
$$
\Pi_{D}=\frac{rcn_{C}}{n}\,.
$$

If the total incentives are used for punishing defectors, individuals who defect have their payoff reduced by $bnu/(n_{D}+1)$, where $b$ characterizes the leverage of punishment \cite{chen_jrsi15} and $n_{D}$ denotes the number of defectors among the other $n-1$ group members. Accordingly, a defector receives
$$
\Pi_{D}=\frac{r c n_{C}}{n}-\frac{bnu}{n_{D}+1},
$$
payoff, while a cooperator in the same group receives
$$
\Pi_{C}=\frac{rc(n_{C}+1)}{n}-c.
$$

In the present work we consider rewarding and punishing in isolation and assume that the total incentives can be only used for rewarding or punishing like Ref.~\cite{sasaki_pnas12}. But differently, we assume that the per capita incentive is not fixed, and can be adjusted during the evolutionary process. We then aim to explore the optimal control protocols of $u$ for institutional reward and punishment respectively. For the sake of proper comparison, we assume that $a=b=1$ in this work.

\subsection{Replicator equation}
We then use the replicator equation to study how the frequencies of different strategies alter in infinitely large, well-mixed populations. Here, we suppose a large population, a fraction $x$ of which encompasses cooperators, and the remaining fraction $1-x$ being defectors. In consequence, the system of replicator equation can be written as \cite{hofbauer98}
\begin{equation}
  \dot{x}=x(1-x)(p_{C}-p_{D}),
\end{equation}
where the average payoffs of cooperators and defectors can be respectively given as
\begin{equation}
p_{i}=\sum_{n_{C}=0}^{n-1}\binom{n-1}{n_{C}} x^{n_{C}}(1-x)^{n_{D}}\Pi_{i}, (i= C \; {\rm or} \; D).
\end{equation}

When institutional reward is used, the average payoffs of cooperators and defectors can be respectively written as \cite{sasaki_dga14,sasaki_pnas12}
$$
p_{C} = \frac{rc[1+(n-1)x]}{n}-c+u\frac{1-(1-x)^{n}}{x},
$$
and
$$
p_{D} =\frac{rc(n-1)x}{n}.
$$
Accordingly, the replicator equation becomes
\begin{equation}
\dot{x}=x(1-x)[u\frac{1-(1-x)^{n}}{x}-\frac{(n-r)c}{n}] \label{eq1}.
\end{equation}

Whereas when institutional punishment is used, the average payoffs of cooperators and defectors can be respectively written as \cite{sasaki_dga14,sasaki_pnas12}
$$
  p_{C}=\frac{rc[1+(n-1)x]}{n}-c,
$$
and
$$
  p_{D}=\frac{rcx(n-1)}{n}-u\frac{1-x^{n}}{1-x}.
$$
Accordingly, the replicator equation becomes
\begin{eqnarray}
\dot{x} = x(1-x)[u\frac{1-x^{n}}{1-x}-\frac{(n-r)c}{n}] \label{eq2}.
\end{eqnarray}

\subsection{Optimal control problems of incentive strategies}

In theory, cooperative behavior will occur as long as the amount of incentives outweights the payoff difference between cooperators and defectors, no matter whether incentive is positive or negative. But implementing incentives is costly and thus we aim to explore the optimal control protocols of $u$ with the minimal amount of cumulative cost for implementing incentives. To do that, we first define a cost function as the objective functional, which is given as
\begin{equation}
J=\int^{t_{f}}_{t_0}\frac{(nu)^{2}}{2}dt, \label{eq3}
\end{equation}
where $t_0$ is the initial time and $t_f$ is the terminal time for the system. In our work, the initial time is $t_0=0$. Based on the above definition, we can then explore the optimal incentive strategies during the evolutionary period between $0$ and $t_f$ using optimal control theory \cite{evans05,geering07,lenhart07}. We consider that $t_f$ is not fixed, but the fraction of cooperators at $t_f$ is fixed at $x(t_f)$. Considering the promotion effects of institutional incentives on cooperation, we further assume that $x(t_f)=1-\delta >x_0$, where $\delta$ is a parameter determining the cooperation level at the terminal time satisfying $0\leq \delta<1$, and $x_0$ is the initial cooperation level in the population.

Based on the above description, we now formulate the two following optimal control problems for institutional reward and punishment respectively. For institutional reward, we have
\begin{equation}
\begin{split}
&\min\,\, J=\int^{t_{f}}_{0}\frac{(nu)^{2}}{2} dt,\\
&{\rm s.t.}\quad  \left\{\begin{array}{lc}
\dot{x}=x(1-x)[u\frac{1-(1-x)^{n}}{x}-\frac{(n-r)c}{n}],\\
x(0)=x_{0}, \\
x(t_{f})=1-\delta.
\end{array}\right. \label{eqa1}
\end{split}
\end{equation}

Analogously, for institutional punishment we have
\begin{equation}
\begin{split}
&\min\,\, J=\int^{t_{f}}_{0}\frac{(nu)^{2}}{2} dt,\\
&{\rm s.t.}\quad  \left\{\begin{array}{lc}
\dot{x}=x(1-x)[u\frac{1-x^{n}}{1-x}-\frac{(n-r)c}{n}],\\
x(0)=x_{0},\\
x(t_{f})=1-\delta.
\end{array}\right. \label{eqa2}
\end{split}
\end{equation}
Here the cost function $J$ characterizes the cumulative cost of one interaction group on average during the period $[0, t_f]$ for the dynamical system described by the replicator equation starting from the initial state $x_0$ to the terminal state $1-\delta$. Thus the quantity $\min J$ can work as the performance index of calculating the optimal control protocol of $u$ with the minimal cumulative cost in a well-mixed population. In what follows, we will use optimal control theory, especially the approach of HJB equation \cite{evans05,geering07,lenhart07}, to solve the two optimal control problems. Specifically, we aim to study whether there exist the optimal control protocols of $u$ which can make the cost function minimized and what their expressions are if exist.

\section{Results}

\subsection{Optimal control strategy of positive incentive}

Using the approach of HJB equation, we can theoretically obtain the optimal control strategy for institutional reward $u_R^{*}(t)$ as (for more detailed theoretical calculations see Appendix A)
\begin{equation}
u_R^{*}(t)=\frac{2(n-r)c}{n}\frac{x}{1-(1-x)^{n}} \label{eq4}.
\end{equation}

Accordingly, with the optimal control strategy $u_R^{*}(t)$, the dynamical system described by Eq.~(\ref{eq1}) becomes
\begin{equation}
\dot{x}=\frac{(n-r)c}{n}x(1-x) \label{eq5},
\end{equation}
where the initial condition is $x(0)=x_0$.

Solving the above differential equation, we have
\begin{equation}
x(t)=\frac{1}{1+(\frac{1}{x_{0}}-1)e^{-\frac{(n-r)c}{n}t}}.
\end{equation}

\begin{figure}
\centering
\includegraphics[width=3.5in]{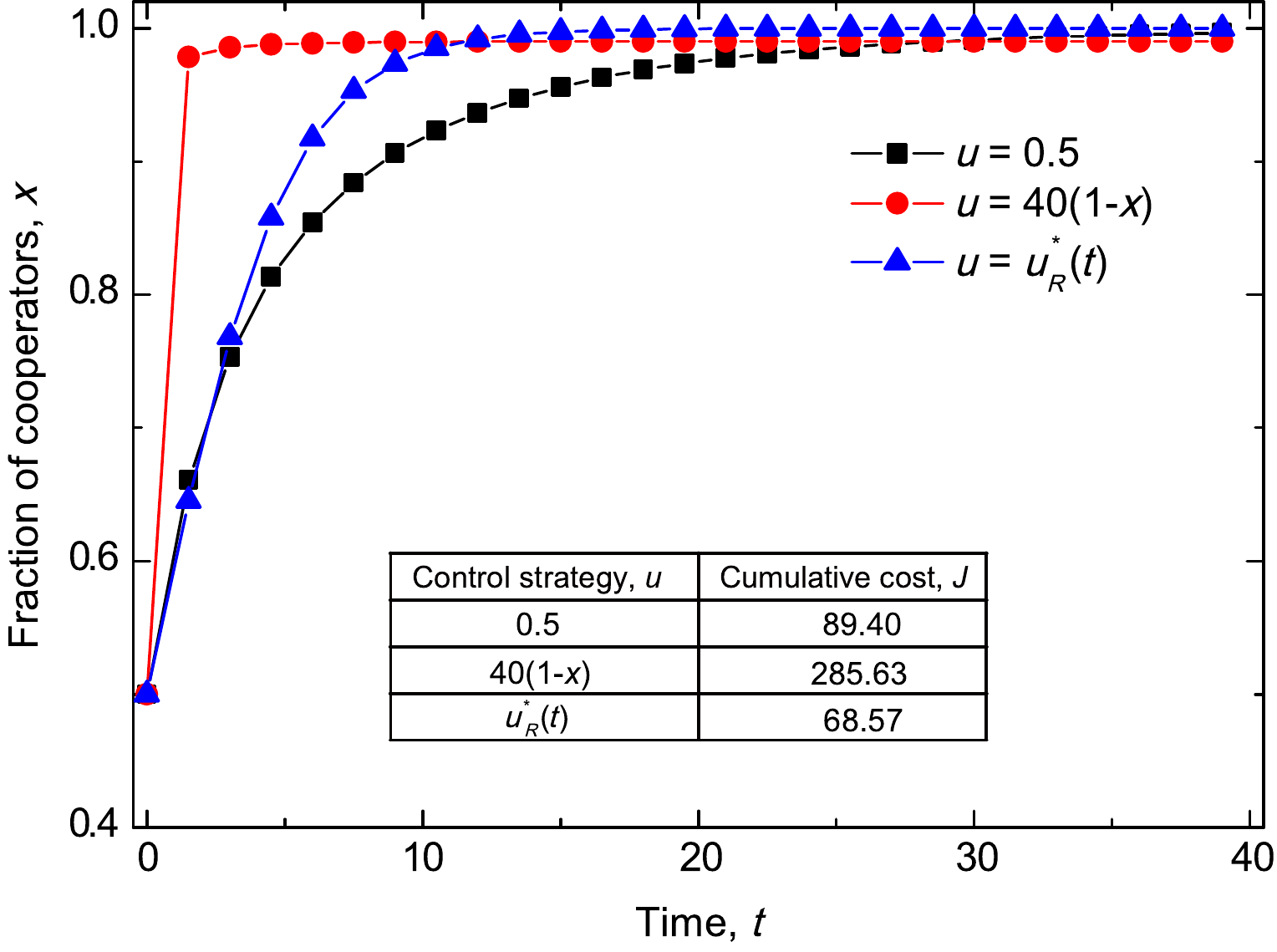}
\caption{Time evolution of the fraction of cooperators for three different rewarding strategies including $u_R^{*}(t)$,  $u(t)=0.5$, and $u(t)=40(1-x)$. The inset table shows the amounts of cumulative cost for the three different rewarding strategies. Parameters: $n=5$, $r=3$, $c=1$, $\delta=0.01$, and $x_{0}=0.5$.}
\label{fig1}
\end{figure}

In order to confirm the above theoretical results for $u_R^{*}(t)$, we would like to provide some numerical examples in what follows. Before that, we point out how to compute the amount of cumulative cost for a given control protocol of $u$. When the initial state and the terminal state are given, we first determine the value of $t_f$ based on Eq.~(\ref{eq1}). We can then obtain the cumulative cost amount using Eq.~(\ref{eq3}). However, when $\delta$ is set to zero, that is, when the terminal state is $x(t_f)=1$, we find that for some given control protocols of $u$ the system needs to take an infinitely long time to reach the terminal state $x(t_f)=1$. In this case, we determine the value of $t_f$ by considering the $\delta$ value is sufficiently small, i.e., $\delta \rightarrow 0$.

In Fig.~\ref{fig1}, we show the fraction of cooperators as a function of time for the obtained optimal rewarding strategy $u_{R}^{*}(t)$ and two other given rewarding strategies $u(t)=0.5$ and $u(t)=40(1-x)$. We find that all these three rewarding strategies can make the dynamical system reach the terminal state and eventually converge to the full cooperation state. But the optimal $u_{R}^{*}(t)$ does not lead to the fastest increase of the fraction of cooperators at the early stage of evolution. Notably, from the inset table of Fig.~\ref{fig1} we find that the amount of cumulative cost for the system to reach the expected terminal state $x(t_f)=0.99$ from the initial state $x_0=0.5$ is about $68.57$ for $u_{R}^{*}(t)$, which is lowest when it is compared with the other two cost values. Thus our numerical calculations support our theoretical results and confirm that the obtained control protocol $u_{R}^{*}(t)$ is superior and dominates other control protocols in the cost of implementing positive incentives.

\subsection{Optimal control strategy of negative incentive}

Similarly, using the approach of HJB equation we can theoretically obtain the optimal control strategy for institutional punishment $u_P^{*}(t)$ as (for more detailed theoretical calculations please see Appendix B)
\begin{equation}
u_P^{*}(t)=\frac{2(n-r)c}{n}\frac{(1-x)}{1-x^{n}}.\label{eq6}
\end{equation}

Accordingly, with the optimal control strategy $u_P^{*}(t)$, the dynamical system described by Eq.~(\ref{eq2}) becomes
\begin{equation}
\dot{x} =\frac{(n-r)c}{n}x(1-x), \label{eq7}
\end{equation}
where the initial condition is $x(0)=x_0$.

Solving the above differential equation, we have
\begin{equation}
x(t)=\frac{1}{1+(\frac{1}{x_{0}}-1)e^{-\frac{(n-r)c}{n}t}}.
\end{equation}

\begin{figure}[!t]
\centering
\includegraphics[width=3.5 in]{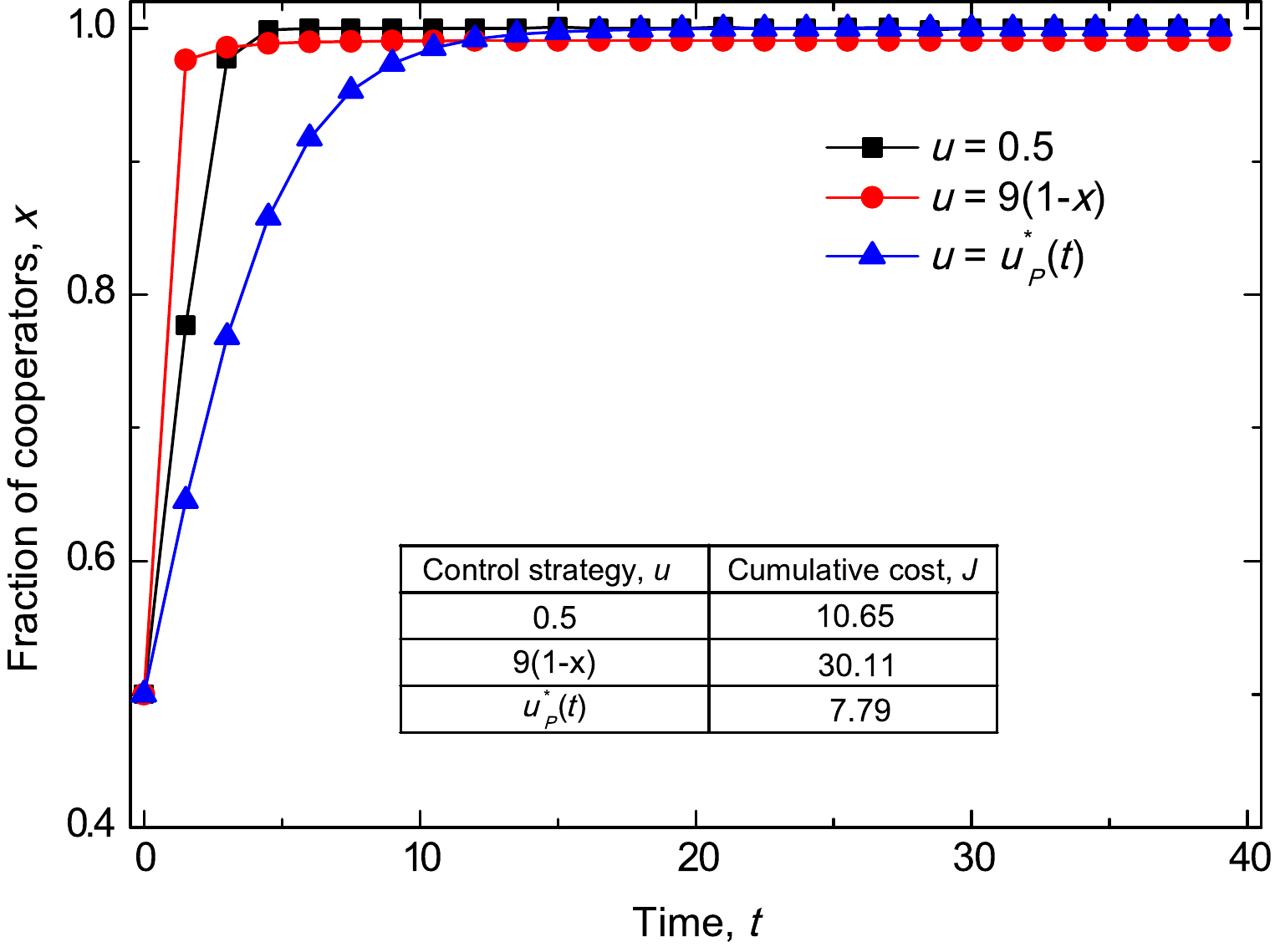}
\caption{Time evolution of the fraction of cooperators for three different punishing strategies including $u_P^{*}(t)$, $u(t)=0.5$, and $u(t)=9(1-x)$. The inset table shows the amounts of cumulative cost for the three different punishing strategies. Parameters: $n=5$, $r=3$, $c=1$, $\delta=0.01$, and $x_{0}=0.5$.}
\label{fig2}
\end{figure}

In order to confirm the above theoretical results, we present some numerical calculations as shown in Fig.~\ref{fig2}. We show that the fraction of cooperators as a function of time for the obtained optimal punishing strategy $u_{P}^{*}(t)$ and two other given punishing strategies $u(t)=0.5$ and $u(t)=9(1-x)$. We find that all these three punishing strategies can make the dynamical system reach the desired terminal state and eventually converge to the full cooperation state, but the optimal $u_{P}^{*}(t)$ makes the slowest increase in cooperation level at the early stage of evolution among the three protocols. Noticeably, we still find that the amount of cumulation cost for the system to reach the terminal state $x(t_f)=0.99$ from the initial state $x_0=0.5$ for the strategy $u_{P}^{*}(t)$ is about $7.79$, which is lower than the other two cost values, as illustrated by the inset table of Fig.~\ref{fig2}. Thus our numerical calculations confirm that the obtained control protocol $u_{P}^{*}(t)$ is better than other control protocols in the cost of implementing negative incentives and support our theoretical predictions.

\subsection{Comparison between optimal positive and negative incentive strategies}

\begin{figure}
\centering
\includegraphics[width=3.5in]{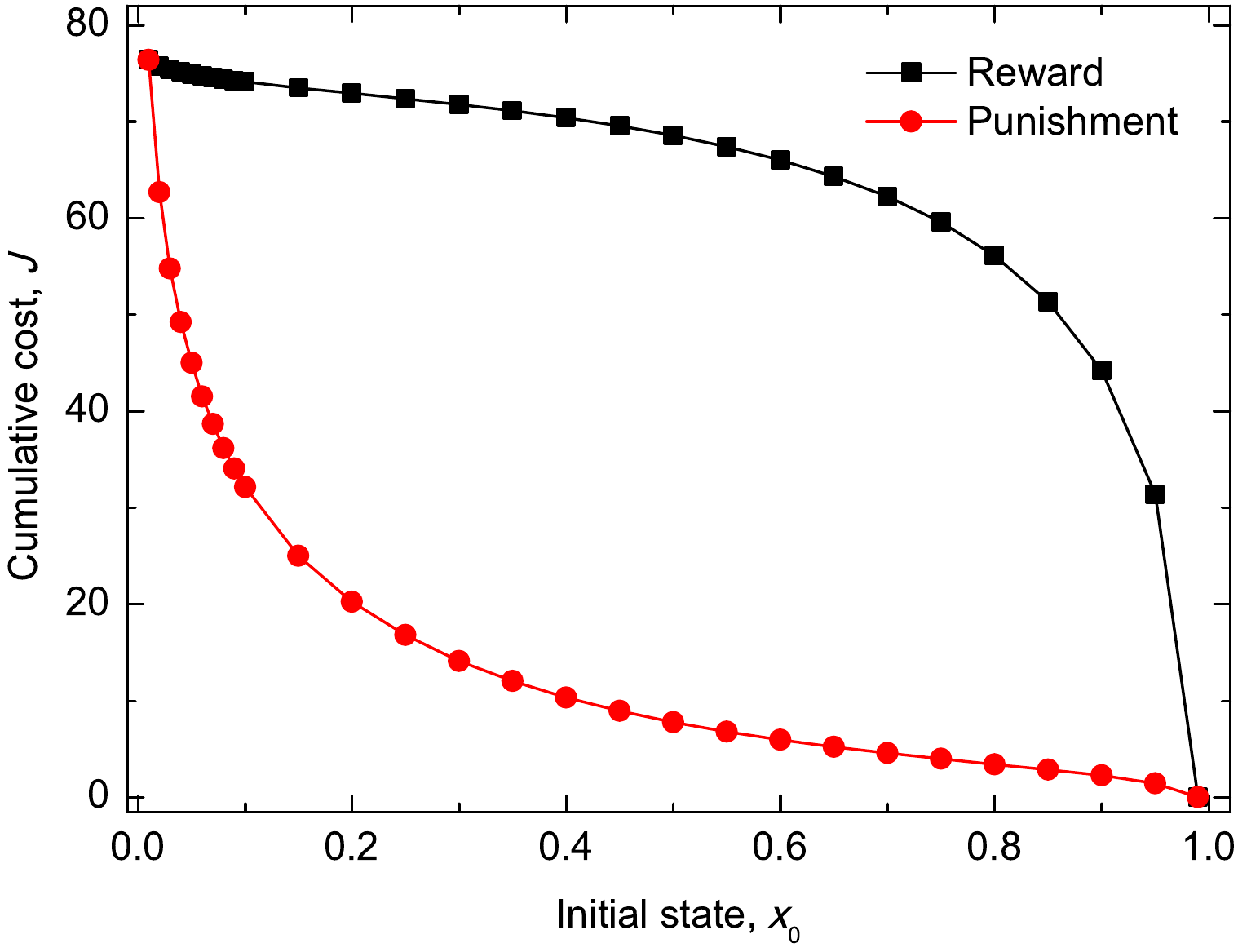}
\caption{The required cumulative cost as a function of the initial state $x_0$ for the optimal rewarding and punishing strategies. Parameters: $n=5$, $r=3$, $c=1$, and $\delta=0.01$.}
\label{fig3}
\end{figure}

Finally, we are interested in making a comparison between the optimal rewarding strategy $u_{R}^{*}(t)$ and the optimal punishing strategy $u_{P}^{*}(t)$. Surprisingly and interestingly, we find that no matter whether the optimal strategy $u_{R}^{*}(t)$ or $u_{P}^{*}(t)$ is used, the dynamical system will become the identical equation $\dot{x} =c(n-r)x(1-x)/n$. This means that both the optimal rewarding and punishing strategies can not only make the dynamical system converge to the full cooperation state, but also have the same value of $t_f$ and the same evolutionary trajectory if the model parameter values are identical. However, due to the expression difference of $u_{R}^{*}(t)$ and $u_{P}^{*}(t)$ we can predict that the two cumulative cost amounts should be different.

In order to have a clear distinction between the two cumulative cost amounts in different conditions, we thus show the cumulative cost values as a function of the initial cooperation level for $u_{R}^{*}(t)$ and $u_{P}^{*}(t)$, as presented in Fig.~\ref{fig3}. We can observe that the cumulative cost monotonically decreases with increasing the initial cooperation level for both optimal positive and negative incentive strategies. In addition, we find that the cumulative cost value of the optimal punishing strategy is lower than that of the optimal rewarding strategy for almost all the $x_0$ values. This indicates that both the optimal rewarding and punishing strategies can make the system reach the same terminal state from the same initial state, but the latter requires less cost. From this viewpoint, the optimal punishing strategy is better than the optimal rewarding strategy.

\section{Discussion}

Prosocial incentive can always promote cooperation, but implementing incentive is costly \cite{hauert_s07,rand_s09,cressman_jtb12,milinski_jtb12}. From the viewpoint of optimization, institutions prefer to adopt the incentive strategy which is able to not only promote cooperation, but also require a low cost \cite{chen_jrsi15,sasaki_pnas12}. In this work, by combining  optimal control theory and evolutionary game approach, we have thus studied two optimal control problems of institutional incentives in a well-mixed population to explore the optimal rewarding and punishing strategies  respectively. By using the method of HJB equation, we theoretically obtain the optimal rewarding and punishing strategies respectively, which can make the cumulative cost minimized and make the dynamical system reach the desired terminal state. Additionally, we provide numerical examples to respectively compare the obtained optimal incentive strategies with other given incentive ones which further confirm our theoretical results. Furthermore, when we compare the optimal rewarding strategy with the optimal punishing strategy, we find that the cumulative cost value for the former is larger than that for the latter in almost all the possible initial conditions by numerical calculations.

In this work, the dynamical system described by the replicator equation is nonlinear, thus it is not easy to obtain the exact expression of the optimal control protocols for institutional incentives in theory \cite{griffin_pre17,bravetti_sr18,wang_ams18}. However, by employing the approach of HJB equation for continuous-time systems, we can succeed in theoretically obtaining the optimal control strategies for institutional incentives. To our best knowledge this is the first time when this dynamic programming method \cite{evans05,geering07,lenhart07} is used in the framework of evolutionary game system. Additionally, these theoretical results are verified by numerical examples.

We find that both the optimal rewarding and punishing strategies can make the dynamical system have the same evolutionary trajectory eventually converging to the full cooperation state, but the former requires less cumulative cost than the latter. Indeed previous works reveal that institutional punishment is a cheaper way of leading the system to reaching the desired cooperation level than applying reward \cite{sasaki_pnas12}. On the one hand, our work thus further confirms this conclusion. On the other hand, however, our work establishes a quantitative index characterizing how much cumulative cost is needed for a given incentive strategy to make the system reach the expected evolutionary destination. Thus, we can quantify how cheaper the optimal punishing strategy is than the optimal rewarding strategy.

A previous work shows that prosocial punishment promotes cooperation, but it does not increase the average payoff of individuals and even reduces the average payoff in some cases \cite{dreber_n08}. Thus people gaining high payoff prefer not to use punishment from the viewpoint of individual-level internal interactions. But institutions, as a top-down-like control mechanism, may not tend to enforce rewarding or punishing on individuals according to individual-level internal interactions. Instead, since institutions need to pay for providing incentives, policy-makers may prefer institutions which not only promote cooperative behaviors, but also are much cheaper \cite{gachter_n12}. Thus our work provides clear arguments why the use of punishment is preferred in human society.

For the above described reason our work only considers the optimal control problems to explore the optimal control strategies with the minimal cumulative cost. Indeed, the time for the system to reach the desired state is also an important quantity for the governance of the public goods \cite{chen_jrsi15,brozyna_ns18,hauser_ns18}. Thus it could be interesting to consider the performance index of time as the objective functional into the optimal control problems of institutional incentives as a possible extension. Work along this line is in progress.

\section*{Appendix A: Theoretical calculations of optimal rewarding strategy}
\renewcommand{\theequation}{a.1\arabic{equation}}
For solving the optimal control problem described by Eq.~(\ref{eqa1}) to theoretically obtain the optimal positive incentive strategy, we use the approach of HJB equation \cite{evans05,geering07,lenhart07}. To begin, we first define the function $f(x,u,t)$ as
\begin{equation}
 f(x, u, t)=x(1-x)[u\frac{1-(1-x)^{n}}{x}-\frac{(n-r)c}{n}].\tag{A1}
\end{equation}
Here $f(x,u,t)$ is a continuously differentiable function coming from the right-hand part of Eq.~(\ref{eq1}) and $u$ is the control variable about positive incentive.

Accordingly, we define the Hamiltonian function $H_R(x,u,t)$ for the equation system as
\begin{align}
H_R(x,u,t)&=\frac{(nu)^{2}}{2}+\frac{\partial J^*}{\partial x}f(x,u,t)\nonumber\\
&=\frac{(nu)^{2}}{2}+x(1-x) \nonumber\\
&[u\frac{1-(1-x)^{n}}{x}-\frac{(n-r)c}{n}]\frac{\partial J^*}{\partial x},\tag{A2}
\end{align}
where $J^*(x,t)$ is the optimal cost function of $x$ and $t$ for the rewarding strategy given as $J^*(x,t)=\int^{t_{f}}_{0}\frac{[nu_R^{*}(t)]^{2}}{2}dt$.

Solving $\frac{\partial H_R}{\partial u}=0$, we know that the optimal positive incentive $u_R^{*}(t)$ should satisfy
\begin{equation}
  u_R^{*}(t) =-\frac{(1-x)[1-(1-x)^{n}]}{n^{2}}\frac{\partial J^*}{\partial x}.\tag{A3}
\end{equation}

In general, in order to obtain the optimal control protocol, we need to solve the nonlinear canonical equations \cite{evans05,geering07,lenhart07}. However, it is quite difficult to solve the equations directly. Instead, we use the dynamic programming method, HJB equation for continuous-time systems \cite{evans05,geering07,lenhart07}, to solve the optimal control problem. Consequently, the HJB equation for the dynamical system with positive incentive can be written as
$$
-\frac{\partial J^*}{\partial t}=H_R[x,u_R^{*}(t),t].
$$

By substituting Eqs.~(A2) and (A3) into the above HJB equation, we have
\begin{align}
-\frac{\partial J^{*}}{\partial t}&= -\frac{(1-x)^{2}[1-(1-x)^{n}]^{2}}{2n^{2}}(\frac{\partial J^{*}}{\partial x})^{2}\nonumber\\
&-\frac{(n-r)c}{n} x(1-x)\frac{\partial J^{*}}{\partial x}.\tag{A4}
\end{align}

Since we assume that the terminal time $t_{f}$ is free, the optimal cost function $J^{*}(x,t)$ is independent of $t$. Consequently, we have
\begin{equation}
  \frac{\partial J^*}{\partial t}=0.\tag{A5}
\end{equation}
We then yield from Eq.~(A4)
 \begin{equation}
 \begin{array}{l}
\frac{\partial J^{*}}{\partial x} = 0 \;\; {\rm or} \;\;
\frac{\partial J^{*}}{\partial x}= -\frac{2n(n-r)cx}{(1-x)[1-(1-x)^{n}]^{2}}.\tag{A6}
  \end{array}
\end{equation}

Since the per capita incentive $u>0$ and $x\in(0, 1)$, from Eq.~(A3) we have
\begin{equation}
  \frac{\partial J^{*}}{\partial x}<0. \tag{A7}
\end{equation}
Therefore only $\frac{\partial J^{*}}{\partial x}= -\frac{2n(n-r)cx}{(1-x)[1-(1-x)^{n}]^{2}}$ holds. By substituting this equation into Eq. (A3), we obtain the optimal rewarding strategy $u_R^{*}(t)$ as
\begin{equation}
  u_R^{*}(t)=\frac{2(n-r)c}{n}\frac{x}{1-(1-x)^{n}}. \tag{A8}
\end{equation}

With the optimal rewarding strategy $u_R^{*}(t)$, the dynamical system thus becomes
\begin{equation}
\dot{x}=\frac{(n-r)c}{n}x(1-x),\tag{A9}
\end{equation}
where the initial condition is $x(0)=x_0$.

By solving Eq.~(A9), we finally obtain the solution of $x(t)$ with the optimal rewarding strategy as
\begin{equation}
x(t)=\frac{1}{1+(\frac{1}{x_{0}}-1)e^{-\frac{(n-r)c}{n}t}}. \tag{A10}
\end{equation}

\section*{Appendix B: Theoretical calculations of optimal punishing strategy}
We solve the optimal control problem described by Eq.~(\ref{eqa2}) to theoretically obtain the optimal negative incentive strategy by using a similar calculation described in Appendix~A. To do that, we first define the function $g(x,u,t)$ as
\begin{equation}
 g(x,u,t)= x(1-x)[u\frac{1-x^{n}}{1-x}-\frac{(n-r)c}{n}].\tag{B1}
\end{equation}
Here $g(x,u,t)$ is a continuously differentiable function coming from the right-hand part of Eq.~(\ref{eq2}) and $u$ is the control variable about negative incentive.

Accordingly, we define the Hamiltonian function $H_P(x,u,t)$ for the equation system as
\begin{align}
H_P(x,u,t) &=\frac{(nu)^{2}}{2}+\frac{\partial J^*}{\partial x}g[x(t), u(t), t]\nonumber\\
&=\frac{(nu)^{2}}{2}+ux(1-x^{n})\frac{\partial J^{*}}{\partial x}\nonumber\\
&-\frac{(n-r)c}{n}x(1-x)\frac{\partial J^*}{\partial x},\tag{B2}
\end{align}
where $J^*(x,t)$ is the optimal cost function of $x$ and $t$ for the punishing strategy given as $J^*(x,t)=\int^{t_{f}}_{0}\frac{[nu_P^{*}(t)]^{2}}{2}dt$.

Solving $\frac{\partial H_P}{\partial u}=0$, we know that the optimal negative incentive $u_P^{*}(t)$ should satisfy
\begin{equation}
  u_P^{*}(t)=-\frac{x(1-x^{n})}{n^{2}}\frac{\partial J^{*}}{\partial x}.\tag{B3}
\end{equation}

Similarly, we use the approach of HJB equation to obtain $u_P^{*}(t)$ by solving Eq.~(B3). And the HJB equation for the dynamical system with negative incentive can be written as
$$
-\frac{\partial J^*}{\partial t}=H_P[x,u_P^{*}(t),t].
$$

By substituting Eqs.~(B2) and (B3) into the above HJB equation, we have
\begin{equation}
    -\frac{\partial J^{*}}{\partial t}=-\frac{x^{2}(1-x^{n})^{2}}{2n^{2}}(\frac{\partial J^{*}}{\partial x})^{2} - \frac{(n-r)c}{n} x(1-x) \frac{\partial J^{*}}{\partial x}.\tag {B4}
\end{equation}

Since we assume that $t_{f}$ is not fixed, the optimal cost function $J^{*}(x,t)$ is independent of $t$. In other words, we have
\begin{equation}
\frac{\partial J^{*}}{\partial t}=0. \tag {B5}
 \end{equation}
From Eq.~(B4) we then obtain
 \begin{equation}
 \begin{array}{l}
\frac{\partial J^{*}}{\partial x} = 0 \;\; {\rm or} \;\;
  \frac{\partial J^{*}}{\partial x}= -\frac{2nc(n-r)(1-x)}{x(1-x^{n})^{2}}.\tag{B6}
  \end{array}
 \end{equation}

Since the per capita incentive $u>0$ and $x\in(0, 1)$, from Eq.~(B3) we have
 \begin{equation}
\frac{\partial J^{*}}{\partial x}<0. \tag{B7}
\end{equation}
Therefore only $\frac{\partial J^{*}}{\partial x}= -\frac{2nc(n-r)(1-x)}{x(1-x^{n})^{2}}$ holds. If substituting this equation into Eq.~(B3), we obtain the optimal rewarding strategy $u_P^{*}(t)$ as
\begin{equation}
u_P^{*}(t)=\frac{2(n-r)c}{n}\frac{(1-x)}{1-x^{n}}.\tag{B8}
\end{equation}

With the optimal punishing strategy $u_P^{*}(t)$, the dynamical system thus becomes
 \begin{equation}
   \dot{x} =\frac{(n-r)c}{n}x(1-x),\tag{B9}
 \end{equation}
where the initial condition is $x(0)=x_0$.

After solving Eq.~(B9), we finally obtain the solution of $x(t)$ with the optimal punishing strategy as
\begin{equation}
   x(t)=\frac{1}{1+(\frac{1}{x_{0}}-1)e^{-\frac{(n-r)c}{n}t}}.\tag{B10}
\end{equation}

\section*{Acknowledgments}
This research was supported by the National Natural Science Foundation of China (Grant No. 61503062) and the Hungarian National Research Fund (Grant K-120785).

\bibliographystyle{elsarticle-num-names}

\end{document}